\documentclass[aps,
twocolumn,
showpacs,amssymb,groupedaddress]{revtex4}
\usepackage{graphicx}
\usepackage{amsmath}
\usepackage{amsfonts}

\newcommand{\be}{\begin{equation}}
\newcommand{\ee}{\end{equation}}

\newcommand{\no}{\nonumber\\}
\newcommand{\ba}{\begin{eqnarray}}
\newcommand{\ea}{\end{eqnarray}}
\newcommand{\bg}{\begin{multline}}
\newcommand{\eg}{\end{multline}}
\def\gl#1{(\ref{#1})}

\newcommand{\la}[1]{\label{#1}}

\begin{document}

\title{Parity doubling from Weinberg sum rules}

\author{A.A.~Andrianov\footnote{ On leave of absence from V.A. Fock Department of Theoretical Physics, St. Petersburg State University,
 Russia; e-mail: andrianov@ecm.ub.es}}\author{D.~Espriu\footnote{
e-mail: espriu@ecm.ub.es}}
\affiliation{Departament d'Estructura i Constituents de la Mat\`eria and
 ICCUB, Institut de Ci\`encies del Cosmos, Universitat de Barcelona,
 Diagonal 647, 08028 Barcelona, Spain}

\begin{abstract}
We investigate the relation among slopes and intercepts of Regge trajectories for mesons
of a given spin and different parities using large $N_c$ arguments and the matching to perturbative QCD in the deep-Minkowski region. For spin-1 mesons of opposite parities we prove that: 
a) for large and increasing $N_c$, the scale $\Lambda^{(V,A)}$ separating the resonance-dominated  and the 
perturbative-saturated region in the channels $V,A$ grows as  $\sqrt{N_c}$; 
b) to satisfy the Weinberg sum rules the slopes of Regge trajectories for mesons of opposite parities must coincide; 
c) their intercepts may differ and their difference corresponds to the difference between $\Lambda^V$ and $\Lambda^A$. Some arguments indicate that 
this difference should tend to zero as $N_c\to\infty$.
\end{abstract}

\pacs{11.15.Pg,11.30.Er,11.55.Hx,12.38.Aw}

\maketitle

\noindent
{\bf 1.}\quad Recently the issue as to whether radial excitations of mesons with a given spin but of opposite parities become
eventually  degenerate in mass in the large $N_c$ limit has been hotly debated \cite{shif1}--\cite{af4}. Different works have clashed
as to whether chiral symmetry restoration at high energies implies that meson masses asymptotically 
approach each other \cite{shif1}--\cite{rab}, or, on the contrary,  the footprint of chiral symmetry breaking persists for 
arbitrarily high mass mesons (in the large $N_c$ limit) \cite{golper}--\cite{shifvain2}.  Some results
based on AdS/QCD correspondence \cite{ckp}--\cite{mpi} have also questioned  whether the slopes of Regge trajectories 
for mesons of opposite parities should be equal. While this issue has a long history (reviewed recently in \cite{gloz2,kaid,af4}),
the increasing interest has been fuelled by the latest improvements in meson phenomenology \cite{kaid}--\cite{kzait}.
In this letter we study the possibility to relate slopes and intercepts of Regge trajectories for mesons
of a given spin and opposite parities by making a careful use of Weinberg sum rules \cite{weinberg,weinberg+}. 

This  issue has been studied traditionally by considering the $N_c\to \infty$ limit at the outset and using the Operator Product Expansion (OPE) in the 
Euclidean region. The main point
of this letter is that keeping $N_c$ large but finite is useful to keep under theoretical control the ``crossover'' 
between the region of resonance saturation and the high energy region where perturbative QCD is
valid. Also we shall work consistently in Minkowski momentum space throughout, as opposed to previous analysis.

Let us consider the correlators
\ba
&&\Pi^{j}_{\mu\nu} (x,y)= \left\langle T\left(J_{\mu} (x) J_{\nu}(y) \right) \right\rangle \no&&\equiv
(-i)\int\frac{d^4 p}{(2\pi)^4} \exp[ip(y-x)] \Pi^{j}_{\mu\nu} (p^2),\label{corr}\\ &&j = V,\, A; \quad
J_{\mu} = \Big({\bar q}(x) \gamma_\mu q(x),\, {\bar q}(x)\gamma_5 \gamma_\mu  q(x)\Big). \nonumber
\ea
As we keep the chiral limit in the major part of our analysis we do not need to specify the internal symmetry 
group and omit the flavor indices. The color degrees of freedom of the quark fields $\bar q, q$ are also omitted in the notation.

For finite $N_c$ two different, non-overlapping, regions of physics can be clearly identified in the physical (Minkowski) momentum region: a region dominated 
by resonances over a non-resonant background, and a region where perturbative QCD  
is reliable. This clear separation is due to chiral symmetry breaking and confinement in the QCD vacuum, on the one side, 
and to the asymptotic freedom of QCD, on the other side. In addition, there is an intermediate region
where neither resonance dominance or perturbative physics describe well the data; resonances are nearly invisible
in the continuum of multiparticle contributions and perturbation theory becomes unreliable. We note that 
large $N_c$ counting rules indicate that the multiparticle background must disappear in the large $N_c$ limit.

At this point we need to be more definite about how we count resonances and in order to do this  
we introduce some `error bar' in the magnitude of the correlators $|\Delta\Pi/\Pi|\sim \epsilon$. Resonances of (relative)
height lower than $\epsilon$ over the  background will be counted as part of the continuum, whereas those that stand out higher than $\epsilon$ will be retained.
The quantity $\epsilon$ will be universal for both correlators.
For  given  $\epsilon$ and $N_c$ 
 one can find a finite number of visible resonances and establish an upper 
bound  $p^2 \leq \Lambda^2_R$ above which one deals with
continuum generated by intermediate multiparticle states but not resolved into resonances.

From the other end, at high energies one expects quark-hadron duality to hold \cite{svz} and perturbation
theory to provide accurate predictions with a (relative) precision
$\epsilon$ down to a scale
$\Lambda^2_{PT}$. By construction, $\Lambda_{R} < \Lambda_{PT}$. At intermediate values $\Lambda^2_{R} < p^2 < \Lambda^2_{PT}$ 
the nonresonant multihadron picture is adequate.

\smallskip\noindent
{\bf 2.}\quad Now let us increase the number of colors.  According to the usual large $N_c$ counting rules we expect that:
(a) Resonances  become narrower and more distinct showing clearer Breit-Wigner shapes and  
increasing their magnitudes. Their position, on the contrary, are independent of $N_c$ at leading order.\quad
(b) At a given value of $p^2$ the number of possible intermediate multiparticle states 
is fixed, but their coupling constants behave as inverse powers of $N_c$ and consequently the nonresonant hadron 
background at fixed $p^2$ decreases. Then for a fixed value of $\epsilon$ more resonances become visible as we increase the number of colors, 
$
\Lambda_{R}(N_c) \le \Lambda_{R}(N'_c),\quad N_c< N'_c . 
%\label{ineq}
$
\quad
(c) The non resonant background due to multiparticle states decreases. 
This background
is lower at lower values of $p^2$ due to phase space considerations.

Let us imagine drawing a band of (relative) width $\epsilon$ around the perturbative prediction. At low values of $p^2$,  perturbation theory fails of course badly in
describing the two point function because resonances are
totally beyond the scope of perturbation theory. As we move
to larger values of $p^2$, the non-resonant background grows and resonances
become broader due to phase space considerations; eventually they all disappear
within the 'error bar' band, thus merging in a continuum. However, this
continuum does not necessarily agree with the one predicted by perturbation theory if the value of $p^2$ is too low.  As we increase $N_c$, the resonances become more marked and more and more of them become visible at a given value of $\epsilon$ and the value of $(\Lambda_R)^2$ increasing with the number of resonances included. The non-resonant, non-perturbative background decreases furthermore as $N_c$ increases. Correspondingly, at the values of $p^2$ where resonances disappear into the continuum, perturbation theory becomes more and more reliable. 

We shall assume a Regge-like behavior for the radially excited states in 
the different channels. Linearly rising trajectories imply that $\Lambda_R^2$
grows linearly with the number of visible resonances. We shall see below that
large $N_c$ counting rules imply that the number of visible resonances increases linearly with $N_c$, so the region of validity of perturbation theory is
reached rather quickly. Combined with the disappearance of the non-perturbative background at large $N_c$, it is rather clear that for any value
of $\epsilon$ there should be a value of $N_c$ large enough (but still finite) where $\Lambda_R\simeq \Lambda_{PT}$.

If we accept this highly plausible conjecture, we can, with an error bounded by
$\epsilon$, replace $\Pi^j_{\mu\nu}(p^2)$ by $\Pi^{j,PT}_{\mu\nu}(p^2)$ for values
of $p^2$ beyond the last visible resonance. 
Whereas it is clear that none of vector resonances saturating the $VV$ correlator in \eqref{corr} must necessarily coincide in 
 mass with the AA one, the chiral symmetry of QCD guarantees the coincidence up to non-perturbative corrections of order $1/(p^2)^3$
of $VV$ and $AA$ correlators at very high momenta.
In fact, the non-perturbative
corrections, that are suppressed by inverse powers of momenta, are also suppressed by powers of $N_c$ as $p^2 \ge \Lambda_{PT}^2\simeq \Lambda_{R}^2\sim N_c$. Thus, at leading order in $N_c$,  $\Pi^{j,PT}_{\mu\nu}(p^2)$ is actually identical for the $V$ and $A$ channels.

Notice that the OPE, valid in the deep Euclidean region, implies in turn via dispersion relations corrections to $\Pi^{j,PT}_{\mu\nu}(p^2)$ proportional to the four-quark condensate in the physical Minkowski region. On dimensional grounds these corrections are down by a power of $1/(p^2)^3$ and in the kinematic region we are considering are of order $1/N_c$, to be compared with the leading perturbative contribution of order $N_c^2$ (assuming again that $(\Lambda_R^j)^2 \sim N_c$). Thus the difference between the $V$ and $A$ channels is much suppressed in the large $N_c$ limit in the region of transition between resonance domination and perturbation theory.

$\Lambda_R^V$ need not be equal to $\Lambda_R^A$ owing to
chiral symmetry breaking. However, taking into account the previous arguments
we can conservatively assume that the difference is of ${\cal O}(1)$, although it is probably even smaller. Assuming that 
$(\Lambda_R^V)^2 - (\Lambda_R^A)^2\sim {\cal O}(1)$  is sufficient for
our purposes.

The number
of visible resonances need not be strictly the same in both channels either. Let
$N^{V}$ and $N^{A}$ be the numbers of such resonances (visible with precision $\epsilon$) for a given $N_c$. If linear Regge trajectories are appropriate for large meson masses,
$(m^{j}_{n})^2 \simeq (m^{j}_{0})^2 + a^{j}  n ,\, n\gg 1, $ then evidently $ N^{j} \sim (\Lambda^{j}_R)^2/ a^{j} $ and increases with growing  $N_c$.

\smallskip\noindent
{\bf 3.}\quad We shall now make use of Weinberg sum rules \cite{weinberg,weinberg+}. Let us decompose the correlators in spin zero and spin one components
\be
\Pi^{j}_{\mu\nu} (p^2) = \Big(- g_{\mu\nu} +\frac{p_\mu p_\nu}{p^2}\Big) \Pi^{j}_{1} (p^2) +
\frac{p_\mu p_\nu}{p^2} \Pi^{j}_{0} (p^2),
\ee
and then use the spectral representation 
\ba
\Pi^{j}_{1} (p^2) = - \int\limits_0^\infty ds \dfrac{\rho^{j} (s)}{p^2 - s +i\varepsilon},\,
\rho^{j} (s) = \frac{1}{\pi}\mbox{\rm Im} \Pi^{j}_{1} (s) > 0,\ \la{spr}
\ea
where $ \rho^{j} (s)$ is related to the probability of producing particles with invariant mass squared $s$.
As to the longitudinal projection one has to remove a possible massless pole in the vector channel and reproduce the pion pole in the axial one
\ba
&&\Pi^{V}_{0} (0) = -\Pi^{V}_{1} (0) =  \int\limits_0^\infty ds \dfrac{\rho^{V} (s)}{s},\no&& \Pi^{A}_{0} (0)= F_\pi^2 - \Pi^{A}_{1} (0) = F_\pi^2 + \int\limits_0^\infty ds \dfrac{\rho^{A} (s)}{s}. \la{spi}
\ea
The current conservation in \gl{corr} for $x\not=y$ is compatible with constant $ \Pi^{j}_{0} (p^2) = \Pi^{j}_{0} (0)$. 
%Schwinger terms are inessential for our considerations.

It is well known that both the spectral representation \gl{spr} and the spectral integrals \gl{spi} are formal, being UV divergent as the probability $ \rho^{j} (s) $
does not decrease at very large $s$, being eventually saturated by the imaginary part of the perturbative decay amplitude into quarks which increases linearly with $s$. As to the IR pole $1/s$ the absence of other massless particles but the pion and the Adler zeroes in the chiral limit guarantee the IR integrability of $ \rho^{j} (s) $.
Thus the dispersion relations need subtractions of the short-distance singularities. On the other hand,
the Wilson analysis of OPE for correlators in $x$ space allows to locate the  singularities on the light cone
which are perturbative due to asymptotic freedom and equivalent for vector and axial-vector channels. Owing to this fact one can combine the difference of VV and AA correlators to eliminate those singularities and derive two well convergent Weinberg sum rules \cite{weinberg,weinberg+},
\ba
&&\int\limits_0^\infty ds \dfrac{\rho^{V} (s)- \rho^{A} (s)}{s} = F_\pi^2,\label{wsr1}\\&& 
\int\limits_0^\infty ds \Big(\rho^{V} (s)- \rho^{A} (s)\Big) = 0.\label{wsr2}
\ea

We now consider these sum rules for a finite but large value of $N_c$ and
assume that  $\Lambda^{V}_R \ge \Lambda^{A}_R $ (the reverse case can be treated similarly and leads to the same results).
Let us saturate the entire spectral density $ \rho^{V} (s)- \rho^{A} (s)$ by well-separated resonances up to $s = (\Lambda^{A})^2$, by resonances for
$\rho^{V} (s)$ and by perturbation theory for $\rho^{A} (s)$ when $(\Lambda^{A})^2 < s < (\Lambda^{V})^2$, as well as by perturbation theory for $\rho^{V,A} (s)$ when $(\Lambda^{V})^2 \leq s$ with
\be \rho_{PT} (s) = N_c C_0 s,\quad C_0 \equiv \frac{1}{24\pi^2}\Big(1 + \frac{N_c \alpha_s 
}{3\pi} + \ldots\Big). \label{denspt}\ee
As previously indicated, the contribution from the condensates can be safely
neglected if $(\Lambda_R^{(V,A)})^2$ is proportional to $N_c$.
Then the Weinberg sum rules \eqref{wsr1} and \eqref{wsr2} read
\ba
&&\sum\limits_{n=0}^{N^{V}} (F^{V}_n)^2 - \sum\limits_{n=0}^{N^{A}} (F^{A}_n)^2 \no&&   = F_\pi^2 +  N_c C_0 \left((\Lambda^{V})^2 - (\Lambda^{A})^2\right),\label{wsr11}\\
&& \sum\limits_{n=0}^{N^{V}} (F^{V}_n)^2  (m^{V}_{n})^2 - \sum\limits_{n=0}^{N^{A}} (F^{A}_n)^2  (m^{A}_{n})^2 \no&& = \frac12  N_c C_0 \Big((\Lambda^{V})^4 - (\Lambda^{A})^4\Big) ,\label{wsr22}
\ea
where we have used the fact that for separated narrow Breit-Wigner resonances one can calculate their individual contributions
\be
\pi \rho^{j}_{n} (s) = \dfrac{(F^{j}_n m^{j}_{n})^2\ \Gamma^j_n m^{j}_{n}}{\Big(s- (m^{j}_{n})^2\Big)^2 + (\Gamma^j_n m^{j}_{n})^2}, 
\ee
extrapolating the integration to infinity and the result is independent of the width. 
%The perturbative contribution of course cancels between
%the $V$ and $A$ channels in the perturbative region.

\smallskip\noindent
{\bf 4.} 
At larger $N_c$ one observes the narrowing and growing of resonances, but they become progressively less marked at higher values of $(m_n^j)^2$. Resonances become invisible when the resonance width $m^{j}_n \Gamma^{j}_n$ becomes comparable with
the distance between neighbor resonances (see similar arguments in \cite{bsz}). For linear trajectories, in the Regge description of mesons \cite{shif1},  $\Gamma^{j}_n \sim B^j m^{j}_n /N_c.$ 
Thus resonances in a given channel overlap when their widths $m^{j}_n \Gamma^{j}_n$ are equal to the corresponding slopes  $m^{j}_n \Gamma^{j}_n \sim B^j (m^{j}_n)^2 /N_c \sim a^{j}$, {\it i.e.} for $ N^{j} \sim N_c/B^j .$
It corresponds to $ (\Lambda^{j}_{R})^2 \sim N^{j}a^{j} \sim  N_c a^{j} /B^j $, showing that the number of visible resonances in
each channel is proportional to $N_c$  as previously indicated.
The corresponding maxima are given by
\be
\pi \rho^{j}_{n} (s)\Big|_{s =(m^{j}_{n})^2} =  \dfrac{(F^{j}_n)^2 m^{j}_{n}}{\Gamma^j_n},
\label{maxim}\ee
At the point where resonances become
invisible (at a fixed value of $\epsilon$), the spectral density levels off at a value
\be  \pi\rho^{j}_{n} (s)\Big|_{s =(\Lambda^{j}_{R})^2} =  \dfrac{(F^{j}_n)^2 (\Lambda^{j}_{R})^2}{a_j}
\simeq  N_c C_0 (\Lambda^{j}_{PT})^2 . \label{rdens}
 \ee
A more precise quantitative estimation of $(\Lambda^{j}_{R})^2$ for a fixed $\epsilon$
is difficult as  the additive Breit-Wigner description of individual resonances is not reliable when there is substantial overlap. Nevertheless a semi-quantitative estimate can be done: let us determine $N^j$ by demanding that oscillations due to resonances relative to the background be $\sim\epsilon$. The value of the minimum between two adjacent resonances (in the Breit-Wigner approximation) is reached for $s \simeq \frac12\Big((m^{j}_{n-1})^2+(m^{j}_{n})^2\Big)$,  and for $n\simeq N^j$ 
\be
\pi \rho^{j}_{n} (s)\Big|_{n=N^j} \simeq  \dfrac{(F^{j})^2 (N^j)^2\frac{B^j}{N_c}}{\frac14  + \frac{(N^jB^j)^2}{N_c^2}}.
\ee
Comparing this with \eqref{maxim}), we get 
 $N^{j} \simeq N_c/2\sqrt{\epsilon}B_j$.

The Regge model also implies
asymptotically equal decay constants; that is $F^{j}_n \sim  F^{j}, \ n\gg 1$. From the previous arguments $
(F^j)^2\simeq N_c C_0 a_j.$
Assuming that  $(\Lambda^{V})^2 - (\Lambda^{A})^2$ is at most of ${\cal O}(1)$ in the large $N_c$ expansion, the Weinberg  sum rules lead immediately to the conclusion that $F^{V} \simeq  F^{A} $ because otherwise in the first Weinberg sum rule \eqref{wsr11}
\[
\sum\limits_{n=0}^{N^{V}} (F^{V}_n)^2 - \sum\limits_{n=0}^{N^{A}} (F^{A}_n)^2 \sim N_c \Big( (F^{V})^2 -
(F^{A})^2\Big)\sim N_c^2
\]
whereas it should be of ${\cal O}(N_c)$.

Next let us analyze the linear Regge asymptotics for radial excitations. Consider meson states lying on the trajectories asymptotically,  $(m^{j}_{n})^2 \simeq (m^{j}_{0})^2 + a^{j}  n ,\, n\gg 1 $. Then, if  $F^{V,A} \simeq  F$, from the second Weinberg sum rule \eqref{wsr22} we get
$a^{V}\simeq a^{A}\equiv a$, {\it i.e.} the slope of trajectories is universal.
Indeed, let us write $(\Lambda^{j})^2 \simeq N^{j} a^{j} + (\Lambda^{j}_0)^2 $. Then up to terms subleading in $N_c$ 
\ba
&& \Big[(F^{V})^2 \Big(\frac12 N^{V}(N^{V}+1) a^{V} + N^{V}(m^{V}_{0})^2\Big)\no&&- (F^{A})^2 \Big(\frac12 N^{A}(N^{A}+1) a^{A} + N^{A}(m^{A}_{0})^2\Big) \Big]\no && \simeq  \frac12 \Big[(N^{V})^2 (F)^2 (a^{V} - a^{A})\Big]\la{match2}\\
\mbox{vs.}\no &&\frac12 N_c C_0\left[\Big(N^{V}a^{V}+ (\Lambda^{V}_0)^2 \Big)^2 
%\right.\no&&\left.
- \Big(N^{A} a^{A}+ (\Lambda^{A}_0)^2\Big)^2\right]\no &&\simeq \frac12 N_c C_0(N^{V})^2 \Big((a^{V})^2 - (a^{A})^2\Big), \nonumber
\ea
which match each other iff $a^{V}= a^{A}$ for $
(F^j)^2\simeq N_c C_0 a_j.$

Finally, from \gl{match2} and from the relation $F^2 = N_c C_0 a$ it follows that 
\be
(m^{V}_{0})^2 -(m^{A}_{0})^2 \simeq (\Lambda^{V}_0)^2 - (\Lambda^{A}_0)^2 
\ee 
at leading order. Indeed, the second sum rule is  saturated by $N^{V} F^2  \Big((m^{V}_{0})^2 - (m^{A}_{0})^2\Big)\simeq N^{V}  N_c C_0 a \Big((\Lambda^{V}_0)^2 - (\Lambda^{A}_0)^2\Big)$ whereas other terms are finite. Thus a possible finite shift between
mass spectra of mesons with different parities in the large $N_c$ approximation has to be accompanied with the same shift in cutoffs for resonance regions, even though as we have argued we expect this difference to be subleading in $N_c$.  However a deviation from the universality $a^{V}= a^{A}$ may also give a comparable term. 
%$ (N^{V})^2 a^{V} - (N^{A})^2 a^{A}\sim (N^{V})^2 \Delta a/N_c \sim N^{V} \Delta a.$ 

We stress once more that the results are not based on the  numbers of resonances  $N^{V}$, $N^{A} $ or cutoffs $\Lambda^{V}_R$,$\Lambda^{A}_R $ being equal. We do expect however that their difference is subleading in $N_c$ for the reasons given above. 

\smallskip\noindent
{\bf 5.}\quad To conclude: For large $N_c$ the region of transition from the resonance dominated region to the perturbatively dominated one is shrinking. The  cutoffs that separate these two regions then: (a) may not coincide in opposite parity channels exactly; ( b) their values squared grow with $N_c$ and their difference is subdominant in the large $N_c$ expansion. In this case, for linear Regge trajectories: (1) the ratio of asymptotic widths to masses of of resonances is the same in opposite-parity channels; (2) the asymptotics of decay coupling constants  as well as the Regge slopes coincide for opposite parity channels; (3) the cutoffs in opposite parity channels don't coincide, but the Regge slopes are universal, and the Regge trajectory intercepts differ in a gap which is fully determined by the difference in the above cutoffs. Similar arguments can be applied to other pairs of channels with the equal quantum numbers but parity.

\smallskip
%\acknowledgments
We are grateful to S. Afonin and J. Soto for  very useful remarks and to A. Vainshtein for fruitful and stimulating discussion (especially during mountain climb in the Pyrenees). This work is  supported by research grants FPA2007-66665, 2005SGR00564, 2007PIV10046. It is also supported by the Consolider-Ingenio 2010 Program CPAN (CSD2007-00042). We acknowledge the  partial support of the EU RTN networks FLAVIANET and ENRAGE and the Program RNP2.1.1.1112.

\end{document}